\documentclass[12pt]{article}
\usepackage{graphicx}
\begin{document}
\bibliographystyle{num}
\title{Dressed vertices}
\baselineskip=1. \baselineskip

\author{     P. Bo\.{z}ek\footnote{Electronic
address~:
piotr.bozek@ifj.edu.pl}\\
Institute of Nuclear Physics, PL-31-342 Cracow, Poland}

\date{\today}

\maketitle

\vskip .3cm


\vskip .3cm

\begin{abstract}

The  response of a
 correlated nuclear system to an external field 
is discussed.
The Bethe-Salpeter equation for the dressed vertex  is
solved.
The kernel of the integral equation for the vertex is chosen
consistently with the  approximation for the self-energy. 
This 
guarantees the fulfillment of the f-sum rule for the response function.

\end{abstract}

{sum rules, vertex corrections, nuclear matter}
\vskip .4cm

{\bf  21.65.+f, 24.10.Cn, 26.60.+c}

\vskip .4cm

Processes occurring in  dense nuclear matter are modified 
by the effects of the medium. This happens  for neutrino rates
in neutron stars \cite{mf,reddyne,haensel,voskrn,yamada,raffelt,sedr} and for 
particle or photon  emission in hot nuclear matter
\cite{Bozek:1997rv,Bozek:1998ro,eff,Cassing:2000ch,Bertsch:1996ig}.
 In particular the effect
of the off-shell propagation of nucleons in the medium is
important for the subthreshold particle production in heavy ion
collisions. The role of  correlations for the neutrino emission could
be especially
important for processes in hot stars.
Soft particle emission is influenced by
the multiple scattering in the medium, which is equivalent to
including vertex correction modifying the coupling of the external current to
the dressed nucleons \cite{Knoll:1996nz}. In-medium modifications of
the vertex cannot be described only as modification of the coupling
constant.
The three-point function describing
the modified coupling of the external current to the fermions in medium has a
complicated analytical structure and depends on the incoming  
momenta and energies \cite{mahan,evans}. 

 For the case of weak coupling of the external current or of
the produced particles to the nucleons, the problem is equivalent to
the calculation of the response function in the correlated  medium.
For the Hartree-Fock approximation the response
 function can be calculated from the random-phase approximation, which
 however is usually taken at the ring diagrams level only.
The use of advanced approximations for  the description
of the many-particle system with in-medium propagators dressed by the
interaction and
scattering,  raises
 the question of the correct treatment of the vertex corrections
 \cite{blarip}.
It is known that
 the
density response function for the electron gas
in the self-consistent GW approximation \cite{GW}
violates the f-sum rule \cite{schin,henn}.
For a given
approximation for the self-energy the corresponding vertex corrections
can be obtained by solving the Bethe-Salpeter equation with a specific
particle-hole irreducible 
kernel 
\cite{kadBaym,blarip}. 
The actual solution of the integral equation for the in-medium vertex
 is quite complex, a complete solution for dressed propagators was
 given by Kwong and Bonitz \cite{bonitz} from the solution of non-equilibrium
 Kadanoff-Baym equations in an external field. This procedure is
 numerically quite involved. The initial state is given 
by a  non-equilibrium evolution of the quantum transport equations and is not
easy to tune. For a quantum well the equations for the dressed
vertex were solved using equilibrium Green's function technique
taking the most important part of the particle-hole irreducible
kernel
\cite{faleev}.
Below we present the first solution of the Bethe-Salpeter equation
for the dressed vertex in a nuclear system for a general momentum and
energy of the external field coupled to the density using 
 the real-time Green's function formalism.

\begin{figure}
\centering
\includegraphics*[width=0.95\textwidth]{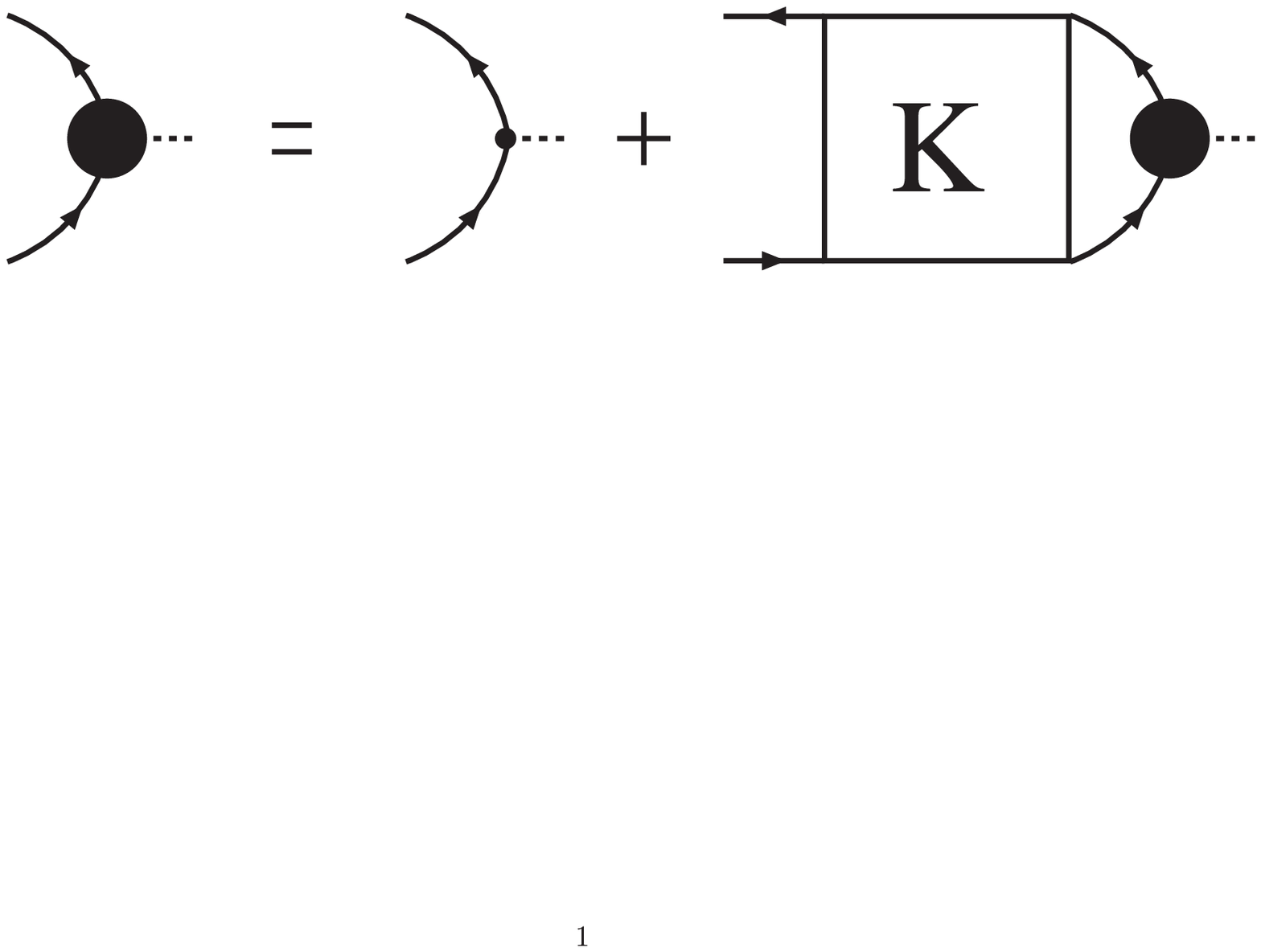}
\caption{The Bethe-Salpeter equation for the dressed vertex. The
  particle-hole
irreducible kernel $K$ is denoted by the box and the fat and the
  small dots denote
  the dressed and the bare vertices respectively. }
\label{bsfig}
\end{figure}

Self-consistent  approximations for  nuclear systems include
iterated second Born approximation using an effective interaction
\cite{Danielewicz:1984ca,Bozek:1998ro,eff} and in medium T-matrix
calculations using free nucleon-nucleon interactions 
\cite{Bozek:2002em,Dewulf:2003nj,Frick:2003sd}.
These schemes use dressed nucleons with nontrivial spectral
functions and  the 
response function cannot be obtained simply by calculating the particle-hole
polarization loop. 
The dressed vertex describing the coupling of the external current to
the in-medium nucleons is given by a Bethe-Salpeter equation
(Fig. \ref{bsfig}), where $K$ denotes the particle-hole irreducible
kernel. For a self-consistent approximation to the self-energy one
should take for the kernel $K$ the functional derivative of the
self-energy with respect to the dressed Green's function 
\cite{kadBaym,blarip} $K={\delta \Sigma}/{\delta G}$.

\begin{figure}
\centering
\includegraphics*[width=0.5\textwidth]{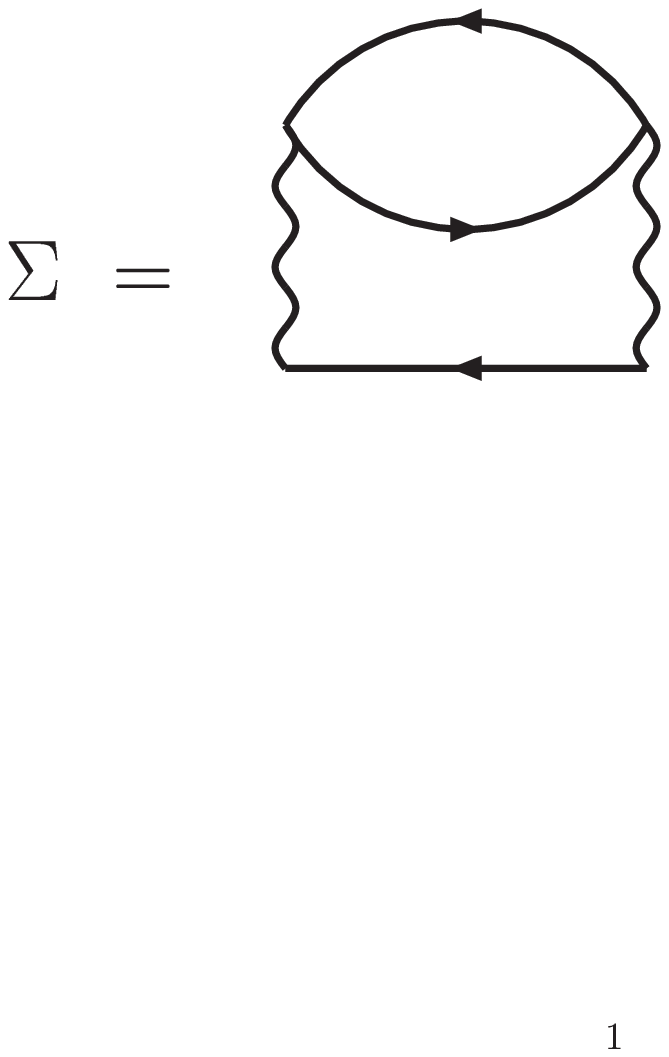}
\caption{The self-energy in the Born approximation.}
\label{selffig}
\end{figure}

In the following we use the self-consistent second Born approximation 
with the effective interaction taken from \cite{Danielewicz:1984ca}
(Fig. \ref{selffig}).
The solution of the self-consistent equations for the self-energy and
dressed propagators in equilibrium is standard 
\cite{Bozek:1998ro,Lehr:2000ua,eff}
and numerical easy to implement.
In the real time formalism the nucleon retarded (advanced) Green's function
is expressed by the retarded (advanced) 
self-energy $\Sigma^{r(a)}$ through the Dyson equation
\begin{equation}
\label{dyson}
G^{r(a)}({\bf p},\omega)=\frac{1}{\omega-{\bf p}^2/2m
-\Sigma^{r(a)}({\bf p},\omega)} \ .
\end{equation}
The self-energy is taken in the second direct Born
approximation
\begin{eqnarray}
\label{self}
\Sigma^{>(<)}({\bf p},\omega)& =& 4 i \int 
\frac{d^3p_1 d\omega_1 d^3p_2 d\omega_2}
{(2 \pi)^8}
V^2({\bf p}-{\bf p_1})G^{>(<)}({\bf p_1},\omega_1)
G^{<(>)}({\bf p_2},\omega_2) \nonumber \\
 & & G^{>(<)}({\bf p}-{\bf p_1}+{\bf p_2},\omega-\omega_1+\omega_2) \ ,
\end{eqnarray}
where the Green's functions
\begin{eqnarray}
G^{>}({\bf p},\omega) &= & -i \left(1-f(\omega)\right)A({\bf
 p},\omega)
 \nonumber \\
G^{<}({\bf p},\omega) &= & i f(\omega)A({\bf p},\omega) 
\end{eqnarray}
are written using the Fermi distribution $f(\omega)$ and
the spectral function
\begin{equation}
\label{spec}
A({\bf p},\omega)=-2 {\rm Im} G^r({\bf p},\omega) \ ,
\end{equation}
also
\begin{equation}
\Sigma^{r(a)}({\bf p},\omega)=
\int\frac{d\omega_1}{2\pi}\frac{\Sigma^<({\bf p},\omega_1)
-\Sigma^>({\bf p},\omega_1)}{\omega-\omega_1\pm i\epsilon} \ .
\end{equation}
Equations (\ref{dyson}), (\ref{self}) and (\ref{spec}) are solved by
iteration with a constraint on the total density of the system.
In the following we present results for the density
$\rho=0.32$fm$^{-3}$ and
the temperature of  $5$MeV. The single-particle width is of about $20-30$MeV.
The polarization bubble without vertex corrections is given by
\begin{equation}
\label{polsimp}
\Pi^{<(>)}({\bf q},\Omega)=- 4 i  \int\frac{d^3p d\omega}{(2\pi)^4}
G^{<(>)}({\bf q}+{\bf p},\omega+\Omega)G^{>(<)}({\bf p},\omega) \ .
\end{equation}
The retarded and advanced polarization $\Pi^{r(a)}$ can be obtained using a
dispersion relation from its imaginary part $2{\rm Im} \Pi^{r}=\Pi^>-\Pi^<$.

\begin{figure}
\centering
\includegraphics*[width=0.95\textwidth]{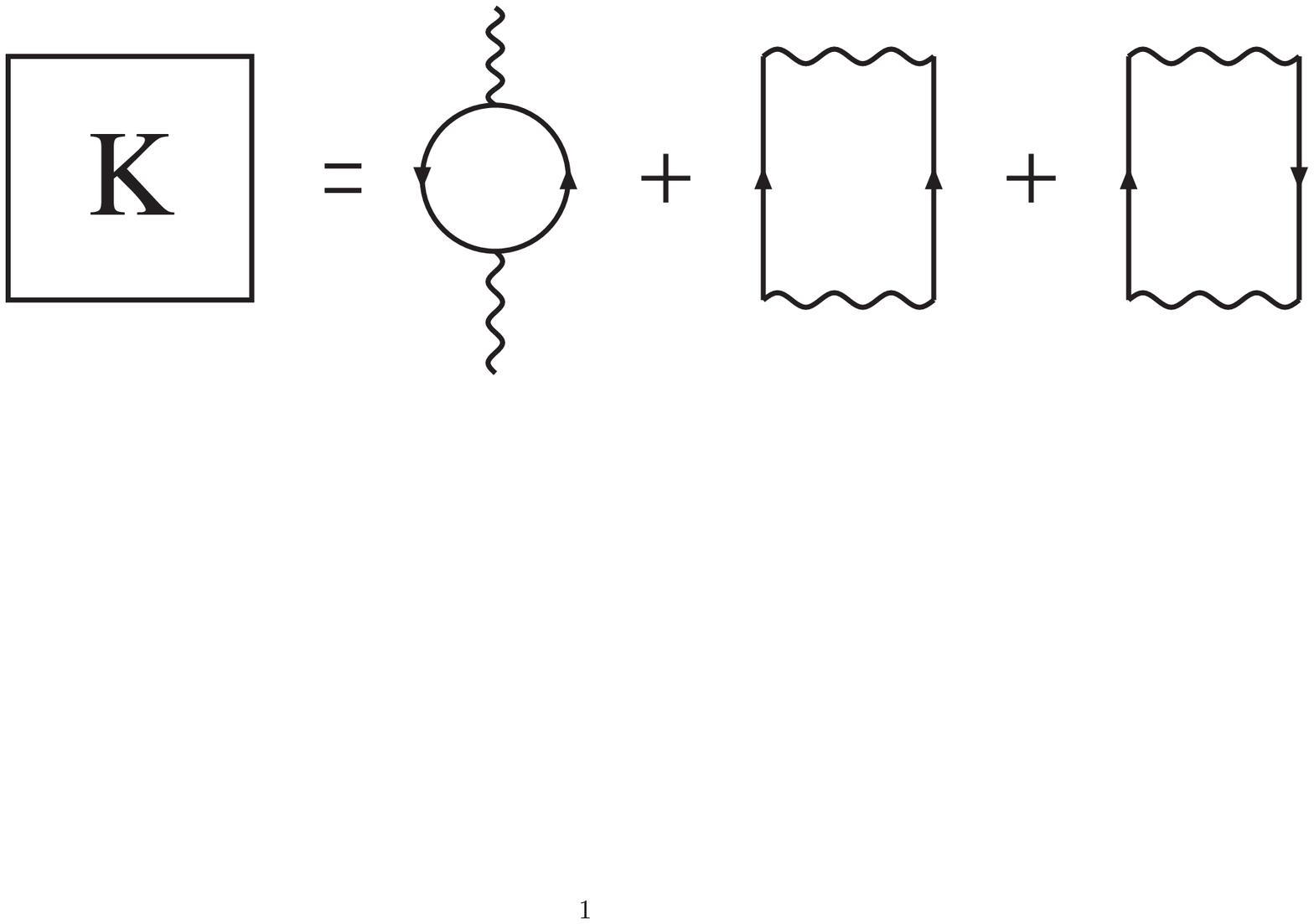}
\caption{The particle-hole irreducible vertex corresponding to the
  self-energy
in Fig. \ref{selffig}.}
\label{kernelfig}
\end{figure}

Let us define the Green's function in a weak external field of
momentum $q$ and energy $\Omega$ coupled to the density.
For the incoming nucleon momentum ${\bf p}$ and energy $\omega$ the
outgoing momentum is ${\bf q}+{\bf p}$ and the energy $\omega
+\Omega$.
We define the smaller  (larger) Green's functions
\begin{equation}
G_{I}^{<(>)}({\bf q}+{\bf p},\omega+\Omega;{\bf p},\omega)
\end{equation}
and the retarded (advanced) Green's functions
\begin{equation}
G_{I}^{r(a)}({\bf q}+{\bf p},\omega+\Omega;{\bf p},\omega) \ ,
\end{equation}
with respect to the ordering of the fermion legs in the three-point
function $G_{I}$, the dependence on the external energy $\Omega$
corresponds always to a retarded vertex.
They fulfill the relation
\begin{equation}
G_{I}^{r}-G_{I}^{a}=G_{I}^{>}-G_{I}^{<}
\end{equation}
but there is no spectral representation for them.
It means that there are three independent Green's functions $G_{I}$
as expected \cite{hou}.
The Green's functions $G_{I}$ are given by the amputated
vertex $\Gamma$ 
\begin{equation}
\label{gdlg}
G_{I}^{r(a)}({\bf q}+{\bf p},\omega+\Omega;{\bf p},\omega)=
G^{r(a)}({\bf q}+{\bf p},\omega+\Omega)\Gamma^{r(a)}({\bf q}+{\bf
  p},\omega+\Omega;{\bf p},\omega)
G^{r(a)}({\bf p},\omega)
\end{equation}
and
\begin{eqnarray}
\label{gdpm}
G_{I}^{<(>)}({\bf q}+{\bf p},\omega+\Omega;{\bf p},\omega)=
G^{r}({\bf q}+{\bf p},\omega+\Omega)\Gamma^{r}({\bf q}+{\bf
  p},\omega+\Omega;{\bf p},\omega)
G^{<(>)}({\bf p},\omega) \nonumber \\
+ G^{r}({\bf q}+{\bf p},\omega+\Omega)\Gamma^{<(>)}({\bf q}+{\bf
  p},\omega+\Omega;{\bf p},\omega)
G^{a}({\bf p},\omega) \nonumber \\
+ G^{<(>)}({\bf q}+{\bf p},\omega+\Omega)\Gamma^{a}({\bf q}+{\bf
  p},\omega+\Omega;{\bf p},\omega)
G^{a}({\bf p},\omega) \ .
\end{eqnarray}
The dressed vertex $\Gamma$ for the coupling of the external field to the
nucleon is the solution of the Bethe-Salpeter equation.
 The particle-hole
irreducible kernel for the self-energy given by  Eq. (\ref{self})
is shown in Fig. \ref{kernelfig}.

\begin{figure}
\centering
\includegraphics*[width=0.7\textwidth]{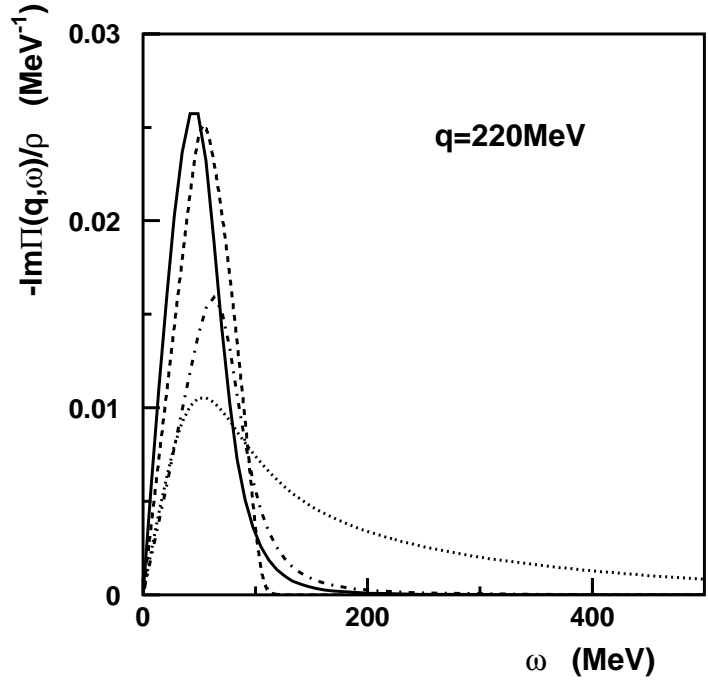}
\caption{The imaginary part of the irreducible polarization 
as function of the energy for $q=220$MeV. The solid line denotes the
result with vertex corrections, the dotted line is the one-loop
calculation with dressed propagators (Eq. \ref{polsimp}), and the
dashed line is the polarization for the noninteracting system.
The dashed-dotted line denotes the polarization with vertex
corrections restricted only to the first diagram in Fig. \ref{kernelfig}.}
\label{polfig}
\end{figure}

Defining the polarization bubble coupled to the external field
\begin{eqnarray}
\Pi_I^{<(>)}({\bf q}+{\bf
  p},\omega+\Omega;{\bf p},\omega)=- 4 i  \int\frac{d^3p_1 d\omega_1}{(2\pi)^4}
\left( G_I^{<(>)}({\bf q}+{\bf
  p}+{\bf p_1},\omega+\Omega+\omega_1;{\bf p}+{\bf
  p_1},\omega+\omega_1)\right. \nonumber \\
\left.
G^{>(<)}({\bf p_1},\omega_1) 
+ G^{<(>)}({\bf p}+{\bf p_1},\omega+\omega_1)G_I^{>(<)}({\bf p_1},\omega_1;
{\bf p_1}-{\bf q},\omega_1-\Omega) 
\right) 
\end{eqnarray}
and
\begin{eqnarray}
\Pi_I^{r(a)}({\bf q}+{\bf
  p},\omega+\Omega;p,\omega)=-4 i  \int\frac{d^3p_1 d\omega_1}{(2\pi)^4}
\left( G_I^{r(a)}({\bf q}+{\bf
  p}+{\bf p_1},\omega+\Omega+\omega_1;{\bf p}+{\bf
  p_1},\omega+\omega_1) \right. \nonumber \\
G^{<(>)}({\bf p_1},\omega_1) 
+ G^{r(a)}({\bf p}+{\bf p_1},\omega+\omega_1)
G_I^{<(>)}({\bf p_1},\omega_1;
{\bf p_1}-{\bf q},\omega_1-\Omega)
\nonumber \\
+ G_I^{<(>)}({\bf q}+{\bf
  p}+{\bf p_1},\omega+\Omega+\omega_1;{\bf p}+{\bf
  p_1},\omega+\omega_1)G^{r(a)}({\bf p_1},\omega_1)
\nonumber \\ \left.
+ G^{<(>)}({\bf p}+{\bf p_1},\omega+\omega_1)
G_I^{r(a)}({\bf p_1},\omega_1;
{\bf p_1}-{\bf q},\omega_1-\Omega) 
\right)
\end{eqnarray}
we can write the Bethe-Salpeter equations for the dressed vertex
 as
\begin{eqnarray}
\label{glg}
\Gamma^{<(>)}({\bf p}+{\bf q},\omega+\Omega;{\bf p},\omega)=
i\int \frac{d^3 p_1 d\omega_1 }{(2\pi)^4} \left( V^2({\bf p}-{\bf p_1})
\Pi^{<(>)}({\bf p}-{\bf p_1},\omega-\omega_1) \right. 
\nonumber \\
G_I^{<(>)}({\bf p_1}+{\bf q},\omega_1+\Omega;{\bf p_1},\omega_1) \nonumber \\
\left.  + V({\bf p_1})V({\bf p_1}+{\bf q})
\Pi_I^{<(>)}({\bf p_1}+{\bf q},\omega_1+\Omega;{\bf p_1},\omega_1)
G^{<(>)}({\bf p}-{\bf p_1},\omega-\omega_1) \right)
\end{eqnarray}
and
\begin{eqnarray}
\label{gpm}
\Gamma^{r(a)}({\bf p}+{\bf q},\omega+\Omega;{\bf p},\omega)=1+
i\int \frac{d^3 p_1 d\omega_1 }{(2\pi)^4} \left( V^2({\bf p}-{\bf p_1})
\Pi^{<}({\bf p}-{\bf p_1},\omega-\omega_1) \right. \nonumber \\
G_I^{r(a)}({\bf p_1}+{\bf q},\omega_1+\Omega;{\bf p_1},\omega_1) \nonumber \\
+V({\bf p_1})V({\bf p_1}+{\bf q})
\Pi_I^{<}({\bf p_1}+{\bf q},\omega_1+\Omega;{\bf p_1},\omega_1)
G^{r(a)}({\bf p}-{\bf p_1},\omega-\omega_1) \nonumber \\
+  V^2({\bf p}-{\bf p_1}) \Pi^{r(a)}({\bf p}-{\bf p_1},\omega-\omega_1)
G_I^{>}({\bf p_1}+{\bf q},\omega_1+\Omega;{\bf p_1},\omega_1)  \nonumber \\
\left. +V({\bf p_1})V({\bf p_1}+{\bf q})
\Pi_I^{r(a)}({\bf p_1}+{\bf q},\omega_1+\Omega;{\bf p_1},\omega_1)
G^{>}({\bf p}-{\bf p_1},\omega-\omega_1) \right) \ .
\end{eqnarray}
The equations for the dressed vertex (\ref{glg}), (\ref{gpm}) and for
Green's function in the external field
(\ref{gdlg}), (\ref{gdpm}) are iterated for each given $q$ and $\Omega$,
using
the Green's functions $G$ obtained for the  correlated system
in equilibrium.
The irreducible polarization bubble including vertex corrections for
the density response is calculated from the Green's function $G_I^<$
\begin{equation}
\Pi_{irr}^{r}({\bf q},\Omega)=-4i\int \frac{d^3p d\omega}{(2\pi)^4}
G_I^{<}({\bf p}+{\bf q},\omega+\Omega;{\bf p},\omega) \ .
\end{equation}
In Fig. \ref{polfig} is show the  polarization bubble
with vertex corrections $\Pi_{irr}$ for $q=220$MeV.
The result is very different from the naive one-loop polarization
 (\ref{polsimp}), the vertex corrections are very important. 
The response function is closer to the response of a noninteracting
system, but not exactly the same. We have made a calculation of the
polarization function neglecting the contributions of the terms with
$\Pi_I$
in the Bethe-Salpeter equations (\ref{glg}), (\ref{gpm}). This corresponds
to neglecting the last two diagrams in the particle-hole kernel 
(Fig. \ref{kernelfig}). The result is different  which shows 
that the full kernel of the Bethe-Salpeter equation
must be taken and not only the first term in Fig. \ref{kernelfig}.

Unlike for the function $G_I$, the real and imaginary parts of the
polarization bubble $\Pi_{irr}^r$ fulfill a dispersion relation
\begin{equation}
\label{disp}
{\rm Re}\Pi_{irr}^r({\bf q},\omega)= -\int \frac{d \omega_1}{\pi}
\frac{{\rm Im}\Pi_{irr}^r({\bf q},\omega_1)}{\omega-\omega_1+ i \epsilon} \ .
\end{equation}
The dispersion relations between the independently
obtained
real and imaginary parts of the response function constitutes a useful
check of the consistency of the approach and
of the numerics.
Another important consistency of the response functions is given by the
f-sum rule
\begin{equation}
-\int \frac{\omega d \omega}{2\pi}{\rm Im}\Pi_{irr}^r({\bf q},\omega)=
\rho \frac{{\bf q}^2}{2m} \ .
\end{equation}
The above sum rule is well satisfied for the free response and for the
response function including vertex corrections, it is severely violated
for the naive polarization loop with dressed propagators (\ref{polsimp}).
Please note that the sum rule is satisfied also by the solution using
a simplified kernel in the Bethe-Salpeter equation, although the
polarization itself is quite different.
Our numerical method is quite accurate for $q>100$MeV, where both the f-sum
rule and the dispersion relation (\ref{disp}) are satisfied. Note that
in the region where the calculations are performed the vertex
corrections are still very important, essential in guaranteeing the
fulfillment of the sum rule. For example at $q=220$MeV the naive
polarization loop with dressed propagators violates the sum rule by a
factor 4.9~.

Adding the Hartree-Fock terms in the self-energy modifies the kernel
of the equation for the dressed vertex. The Fock term generates an
additional interaction line in Fig. \ref{kernelfig}, the exchange term
in the random-phase approximation. The Hartree term generates the ring
series which is summed to give the polarization
\begin{equation}
\label{ring}
\frac{\Pi^r_{irr}({\bf q},\Omega)}{1-V({\bf q})\Pi^r_{irr}({\bf q},\Omega)} \ .
\end{equation}
We have checked that adding at the same time a Hartree-Fock term 
to the self-energy (\ref{self}) and an exchange interaction ladder in
the kernel of the Bethe-Salpeter equation conserves the sum rule
for $\Pi_{irr}^r$. This sum rule is then conserved after performing
the ring summation (\ref{ring}), although the shape of the response
function itself is strongly modified by the transformation (\ref{ring}).
The irreducible response function with vertex correction is close to
the
noninteracting one, however after the ring summation the results are
different due to a different tail in ${\rm Im}\Pi_{irr}$ at large energies.
The study of the full response with a realistic effective Hartree-Fock
self-energy will be presented elsewhere.

We  present a solution of the Bethe-Salpeter equation for the
vertex of the external field coupled to the density in a correlated
medium.
The approximation chosen for the particle-hole interaction 
is consistent with the {\it nontrivial}
self-energy used in the Dyson equation. It guarantees the fulfillment of
the f-sum rule for the response function.
To our knowledge it is the first solution of the Bethe-Salpeter
equation for a general momentum and energy, using equilibrium Green's
function formalism and the full kernel of the integral equation for
the dressed vertex. 
 The irreducible polarization with vertex corrections is much closer
 to the Lindhard function than the naive one loop polarization with
 dressed propagators. It is a manifestation of the expected
 cancellation of the self-energy and vertex corrections.
The methods here  presented can be used to calculate realistic response
functions with different vertices or to test approximations for the dressed 
vertices or for the particle-hole interaction for the more complicated
T-matrix self-energies in nuclear matter.

\bf Acknowledgments
\vskip .3cm
This work was partly supported by the KBN
under Grant No. 2P03B05925.

\bibliography{../mojbib}

\newpage

\begin{figure}
\centering
\includegraphics*[width=0.7\textwidth]{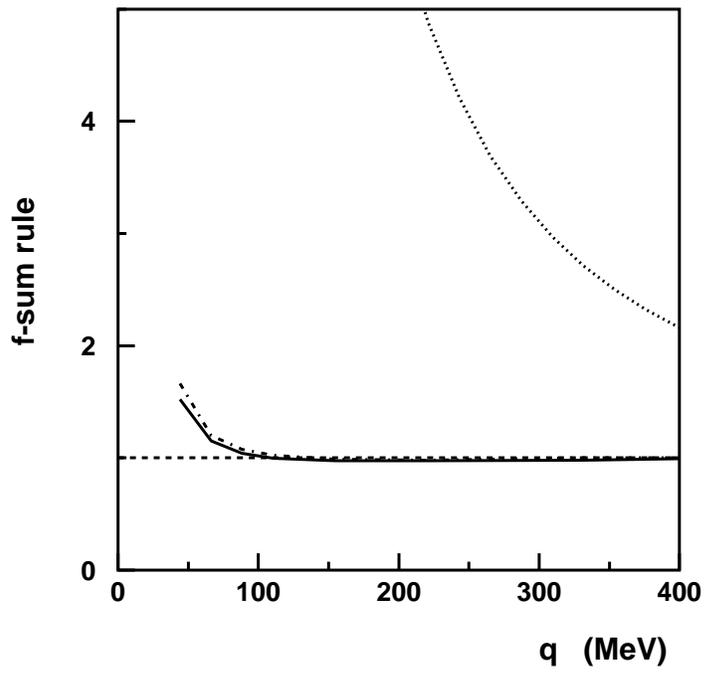}
\caption{The ratio of the f-sum rule to the expected value as function
  of the momentum in the response function. The lines are the same as
  in Fig. \ref{polfig}.}
\label{sumrfig}
\end{figure}

\end{document}